\documentclass[11pt,a4paper]{article}
\usepackage{cite}
\usepackage{a4wide}
\usepackage{amsmath,amssymb,amsfonts}
\usepackage{graphicx}
\usepackage{textcomp}
\usepackage{xcolor}
\usepackage{listings}
\usepackage{todonotes}
\usepackage{hyperref}
\usepackage{enumitem}

\def\BibTeX{{\rm B\kern-.05em{\sc i\kern-.025em b}\kern-.08em
    T\kern-.1667em\lower.7ex\hbox{E}\kern-.125emX}}

\definecolor{keywordcolor}{RGB}{20,105,176}
\definecolor{valuecolor}{RGB}{106, 109, 32}
\definecolor{string}{rgb}{0.64,0.08,0.08}

\lstdefinestyle{FormattedNumber}{%
    literate={0}{{\textcolor{red}{0}}}{1}%
             {1}{{\textcolor{red}{1}}}{1}%
             {2}{{\textcolor{red}{2}}}{1}%
             {3}{{\textcolor{red}{3}}}{1}%
             {4}{{\textcolor{red}{4}}}{1}%
             {5}{{\textcolor{red}{5}}}{1}%
             {6}{{\textcolor{red}{6}}}{1}%
             {7}{{\textcolor{red}{7}}}{1}%
             {8}{{\textcolor{red}{8}}}{1}%
             {9}{{\textcolor{red}{9}}}{1}%
             {?}{{\textcolor{red}{?}}}{1}%
             ,
   basicstyle=\ttfamily,
}

\newcommand{\tac}{{\tt tac}}
\newcommand{\sort}{{\tt sort}}
\newcommand{\gzip}{{\tt gzip}}
\newcommand{\rpc}{{\sc rpc}}
\newcommand{\json}{{\sc json}}
\newcommand{\field}[1]{{\tt #1}}
\newcommand{\myline}[1]{{\small\tt `#1'}}
\newcommand{\Kin}{\ensuremath{K_{\mathit{in}}}}
\newcommand{\Kout}{\ensuremath{K_{\mathit{out}}}}
\newcommand{\kin}{\ensuremath{k_{\mathit{in}}}}
\newcommand{\kout}{\ensuremath{k_{\mathit{out}}}}

\newcommand{\nbblocks}{674,001}
\newcommand{\nbtransactions}{623,483,734}
\newcommand{\nbTIO}{1,673,052,718}

\begin{document}

\title{
{\bf
Full Bitcoin Blockchain Data Made Easy
}\\
\large
\bigskip
{\bf Jules Azad Emery and Matthieu Latapy}\\
\medskip
Sorbonne Universit\'e, CNRS, LIP6, F-75005 Paris, France\\
\smallskip
\url{bitcoin@complexnetworks.fr}
}

\date{}
\maketitle

\vspace*{-1.5cm}

\begin{abstract}
Despite the fact that it is publicly available, collecting and processing the full bitcoin blockchain data is not trivial. Its mere size, history, and other features indeed raise quite specific challenges, that we address in this paper.
The strengths of our approach are the following: it relies on very basic and standard tools, which makes the procedure reliable and easily reproducible; it is a purely lossless procedure ensuring that we catch and preserve all existing data; it provides additional indexing that makes it easy to further process the whole data and select appropriate subsets of it. We present our procedure in details and illustrate its added value on large-scale use cases, like address clustering. We provide an implementation online, as well as the obtained dataset.
\end{abstract}

\section*{Introduction}

Bitcoin \cite{whitepaper} is the first and most widely used crypto-currency. Introduced in 2009, its market capitalization was above 1 trillion U.S. dollars in february 2021. With around 300,000 transactions daily, it is used worldwide in a variety of situations.

{\bf All bitcoin transactions are recorded in a public registry, called the blockchain}. It is a sequence of blocks issued every ten minutes, each listing the transactions successfully processed since the previous block. This gives a unique opportunity to study real-world, large-scale financial activity.
Therefore, bitcoin is at the core of an intense research activity.

However, accessing and processing bitcoin data raises serious challenges. First, the blockchain is ruled by complex protocols that evolved over time, making it necessary to have more than a basic understanding of how it works. In addition, precisely because the blockchain is public, its users use a wealth of obfuscation techniques to preserve their privacy and make transactions anonymous. Last but not least, the sheer size of the data makes it difficult to collect, and even more difficult to analyze; even trivial questions like counting the number of transactions involving a given user are difficult, at such a scale.

As a consequence, most bitcoin studies rely on partial views of the blockchain, on aggregated data, or on privately owned datasets; there is currently no publicly available and easy-to-use dataset containing all bitcoin information stored in the blockchain since its beginning.

{\bf Our goal with this paper is to provide such a dataset.} We want to provide {\em all} existing data, without any assumption on what the final user will need. In addition, we want to pre-process these data to add the information needed to use it efficiently and easily. In order to ensure that our procedures are reliable, efficient, and reproducible, we want to rely uniquely on basic, standard command-line tools.


\noindent
Our approach consists in a sequence of key steps, that we detail in the following sections of this paper:
\begin{enumerate}[label=\Roman*.]
\item {\bf Collection.} We first perform the raw data extraction from the bitcoin blockchain itself, by setting up a bitcoin node and querying it for each block.
\item {\bf Indexing.} We add to the raw data an integer index for each basic item, like transactions or addresses, without removing any of the original data.
\item {\bf Distillation.} We extract specific subsets of the data that make it easy, fast, and compact to obtain various higher-level information.
\item {\bf Application.} In order to illustrate our contribution, we detail an advanced use of our dataset: the analysis of a classical address clustering heuristic.
\end{enumerate}

\noindent
We discuss {\bf related work} in Section~\ref{sec:related}, and {\bf perspectives} in Section~\ref{sec:discussion}.

{\bf All the code presented in this paper and all obtained data are documented and available online \cite{URL}.}
We give approximate execution times of each step, but they strongly depend on available bandwiudth, memory space, computation power, disk speed, implementation language, and other parameters; they should therefore be considered as indicative only.


\section{Collection}
\label{sec:collection}

In order to have a local copy of the blockchain, {\bf we first set up our own bitcoin node}. To this end, we install and run the open-source {\em bitcoin-core} software \cite{bitcoin-core}. It contacts a set of DNS nodes hard-coded in its source code, from which it obtains a list of running bitcoin nodes. Our node then downloads the blockchain from these nodes, thus obtaining its local copy.

We launched this procedure on March 13, 2021. It downloaded the blockchain available at that date in less than $6$~hours, leading to a use of $360$~GB of disk space.

In principle, one may then read and decode the local binary files used by the node to store its copy of blockchain data. This would be the fastest solution, but it is complex and prone to errors. Indeed these file formats are poorly documented, they changed over years, and they may change again in a close future. In addition, this approach only collects targeted parts of the data, in general. See for instance \cite{AbeURL}, and Section~\ref{sec:related} for more details.

We therefore adopted a slower but simpler and safer approach. It is based on \rpc\ ({\em Remote Procedure Call)}, the protocol provided to monitor bitcoin nodes \cite{getblock-rpc}. It implements in particular a primitive that returns the block identifier of the $i$-th block in the chain, for any given $i$; and a primitive that returns a \json\  object containing all the data available in a block of given identifier. Figure~\ref{fig:json} presents a simplified version of such a \json\ object, that we will detail below.

Thanks to these primitives, {\bf we ran an \rpc-based bitcoin blockchain data collection under the form of a \json\ object for each block}. Instead of running both primitives for each block, we took benefit of the fact that each block contains the identifier of the previous block in the chain. We therefore collected the latest block (its number is provided by a query to any bitcoin node) and then iteratively collected the previous block until the beginning of the blockchain is reached.

We launched this data collection until block $674\,000$, dated March 10th, 2021, which took $58$~hours. It produced a $2.1$~TB ($548$~GB once compressed with \gzip) text file in which each line is the \json\ object describing a block. Notice however that obtained blocks are in reverse chronological order (the latest one first), which is not convenient for further analysis. We therefore reversed the initial file using the classical \tac\ tool. It needs a non-compressed file as input, but it has the advantage of not storing the whole file into main memory, which is crucial here. It performed the reversing of the \nbblocks\ lines of the $2.1$~TB file in $28$~hours.

\begin{figure}[!h]
\centering
\begin{minipage}{.65\textwidth}
\input{json_ex.tex}
\end{minipage}
\caption{
{\bf Simplified example of a block description in \json\ format.}
This is block number \field{2}, recorded at time \field{t}, and it contains transactions, \field{B} and \field{E}. Transaction \field{B} has two inputs: the output \field{0} of transaction \field{C}, and the output \field{0} of transaction \field{D}. It has two outputs: its output \field{0} goes to addresses $a$ and $c$; its output \field{1} goes to address \field{c}. Transaction \field{E} has one input: the output \field{1} of transaction \field{B}; and it has two outputs: its output \field{0} to address \field{f} and its output \field{1} to address \field{e}. The \field{value} fields indicate that transaction \field{B} sends \field{2} satoshis to addresses \field{a} and \field{c} and \field{5} satoshis to address \field{c}, taken from outputs \field{0} of transactions \field{C} and \field{D}; and that transaction \field{E} sends \field{3} satoshis to address \field{f} and \field{2} to address \field{e}, taken from output \field{1} of transaction \field{B}. See Figure~\ref{fig:ex-graphical} for a graphical representation.
}
\label{fig:json}
\end{figure}

As illustrated in Figure~\ref{fig:json}, each \json\ description of a block has several fields.
It begins by fields characterizing the block itself, including its rank in the blockchain (\field{height} field), its identifier (a hash code) and the one of its previous and next block, as well as a timestamp. It has another crucial field, named \field{tx}, that gives the list of transactions recorded in this block.

Each transaction in this list is itself described by a \json\ object, with main fields \field{txid}, \field{vin} and \field{vout}. The \field{txid} field is the transaction identifier, that we will represent by a capital letter. Fields \field{vin} and \field{vout} are the lists of this transaction inputs and outputs (TIOs for short), respectively.

Each transaction output is described by a rank field \field{n} and has a \field{value} field giving the amount, in satoshis (a satoshi is $10^{-8}$ bitcoin)\,\footnote{Amounts are actually stored as an integer number of satoshis in the blockchain, but returned as a decimal number of bitcoins by \json\ \rpc\ calls. We convert them into satoshi units to avoid rounding errors.} sent to this output. Addresses may be associated to outputs, described by a \json\ object in field \field{scriptPubKey}. We will represent addresses with lower case letters here.

Each transaction input is the output of a previous transaction; it is described by the \field{txid} of this transaction and the rank of its output under concern, given in a field named \field{vout} too.

Addresses require specific attention, as their format is not uniform. As already said, they are given in the \field{scriptPubKey} \json\ object associated to transaction outputs. This \json\ object has a \field{type} field, and if its value is \field{nulldata} or \field{nonstandard}, then this output has no directly available address. This is rare, though; in most case, either this \json\ object has field named \field{addresses} that gives the addresses under concern, or the \field{type} field has value \field{pubkey}. In this last case, the address is the first word (with space separator) in another field, named \field{asm}. In order to make the data easier to parse, we then add an \field{addresses} field to each \field{scriptPubKey} \json\ object, with the addresses we found in it.

\begin{figure}[!h]
\centering
~\hfill
\begin{minipage}{.3\columnwidth}
\centering
\includegraphics[width=\textwidth]{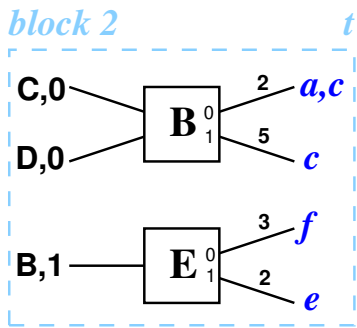}
\end{minipage}
\hfill
\begin{minipage}{.5\columnwidth}
\caption{
{\bf Graphical representation of a bitcoin block and its transactions.}
Their \json\ description is in Figure~\ref{fig:json}. Transactions are represented by square boxes, with their capital letter identifier and output ranks. Addresses are in blue, lower case letters. Amounts appear on output edges of transactions, and the inputs give the corresponding transaction identifier and its output rank.
}
\label{fig:ex-graphical}
\end{minipage}
\hfill~
\end{figure}

{\bf We graphically represent} the key information contained in a block like in Figure~\ref{fig:ex-graphical}. In this figure, we represent the block of Figure~\ref{fig:json}.

Let us insist on the fact that we do not detail {\em all} available fields above, and we do not represent them all on the picture. There is a wide variety of transaction types and other specific features, that vary over time. All these data are present in our dataset, but they are not our focus here.

With these data, {\bf we already obtain some basic but interesting statistics}. In particular, the dataset contains \nbblocks\ blocks and \nbtransactions\ transactions with \nbTIO\ inputs and outputs.

\begin{figure}[!h]
~\hfill
\includegraphics[width=.4\columnwidth]{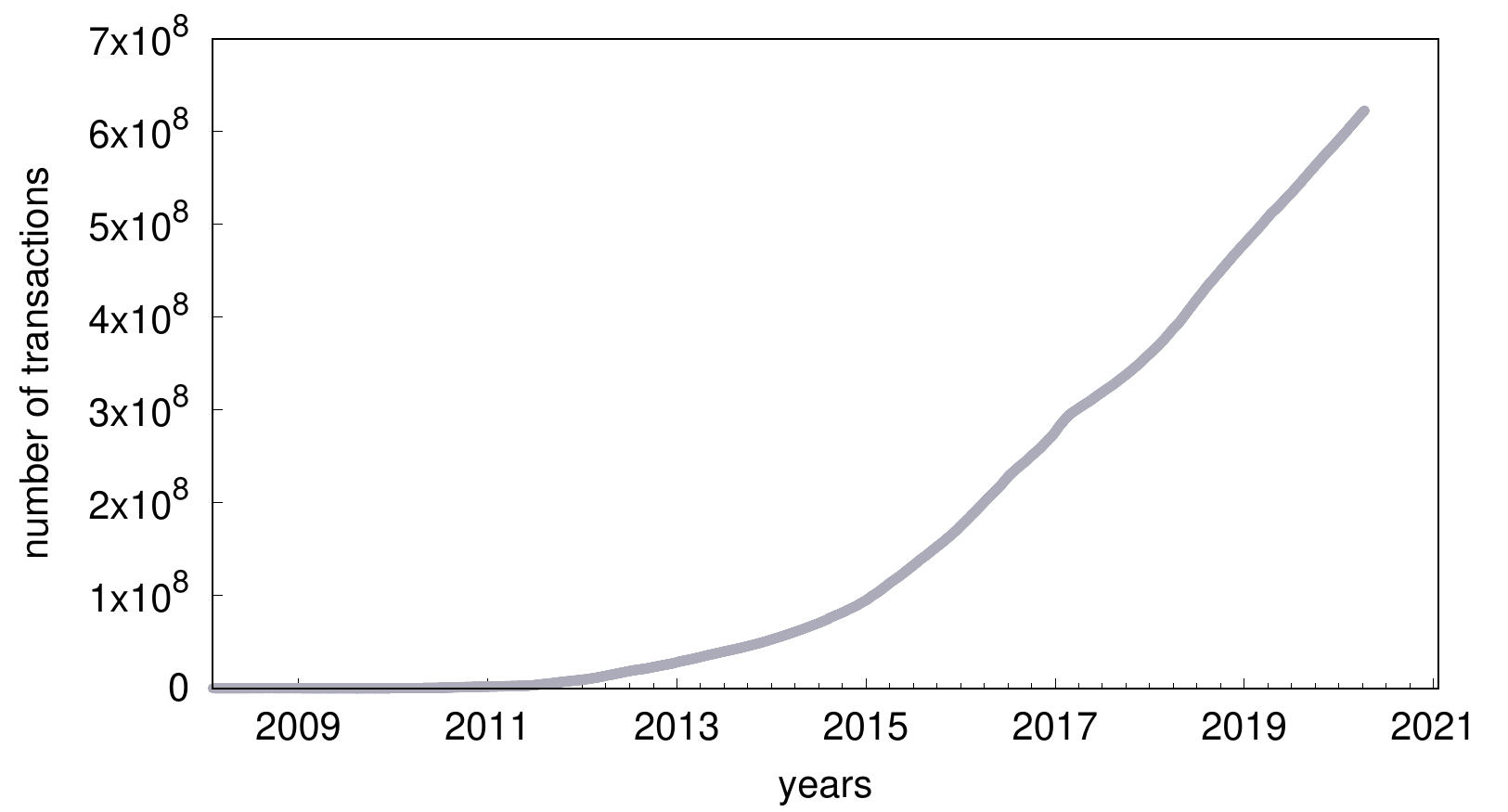}
\hfill
\includegraphics[width=.4\columnwidth]{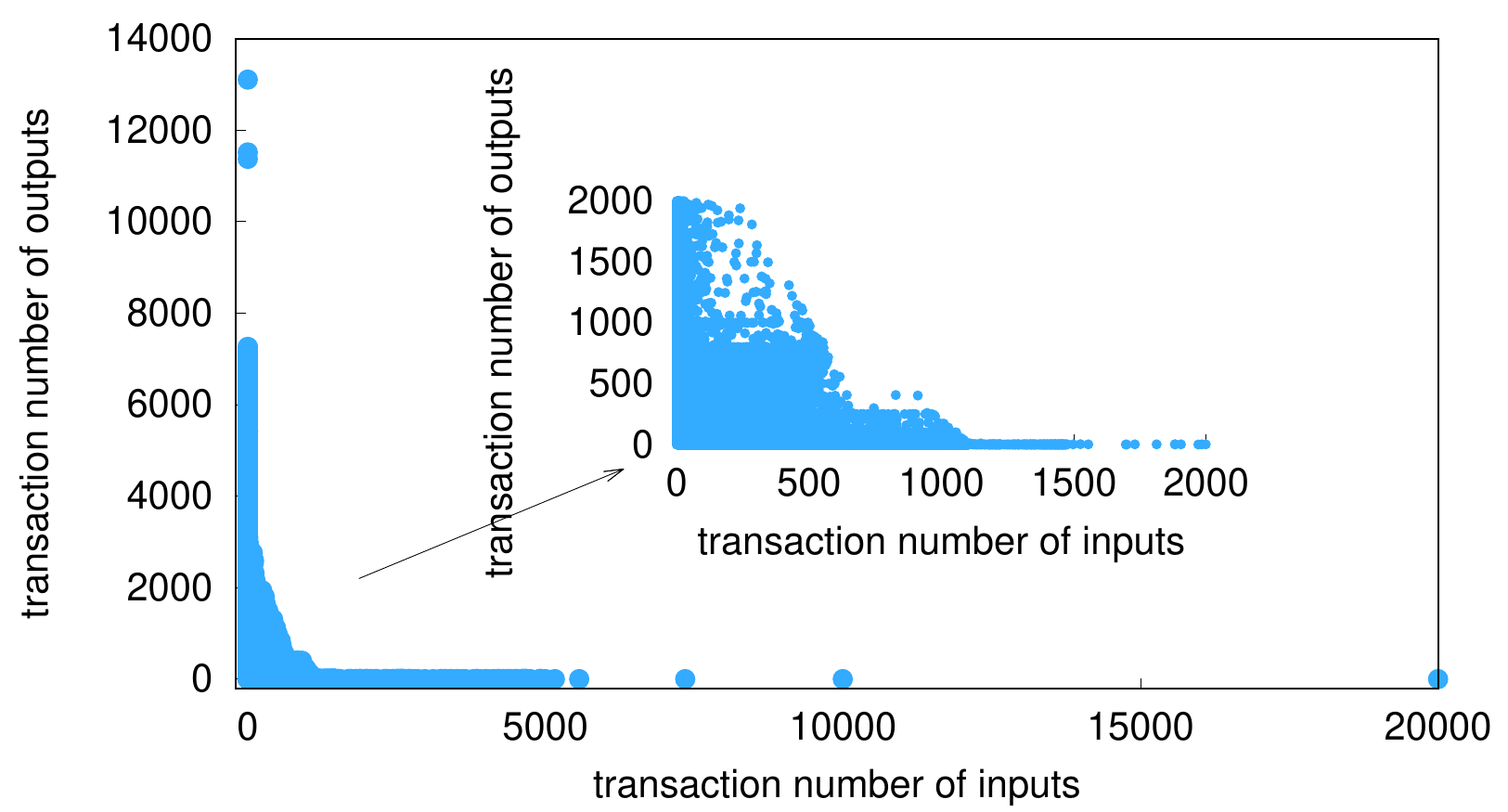}
\hfill~
\caption{
{\bf Left: number of transactions over time}, since the beginning of the bitcoin blockchain.
{\bf Right: correlations between transaction number of inputs and outputs.} For each transaction with $x$ inputs and $y$ outputs, we display a dot at coordinates $(x,y)$. The inset provides a zoom on smallest values.
}
\label{fig:nb_trans_t_and_io_correlations}
\end{figure}

As an illustration, we present in Figure~\ref{fig:nb_trans_t_and_io_correlations}~(left) the number of transactions over time, that displays the classical slow start until 2013, followed by a rapid growth that nowadays becomes linear.
Figure~\ref{fig:nb_trans_t_and_io_correlations}~(right) presents the correlations between transaction number of inputs and outputs. These numbers span orders of magnitude, but large number of inputs are for transactions with only few outputs, and conversely. This indicates specific kinds of transactions, forged for obfuscation, as we will detail in Section~\ref{sec:application}. Instead, most transactions have only few inputs and outputs, and then the values are not significantly correlated (see the inset).

\begin{figure}[!h]
~\hfill
\includegraphics[width=.4\columnwidth]{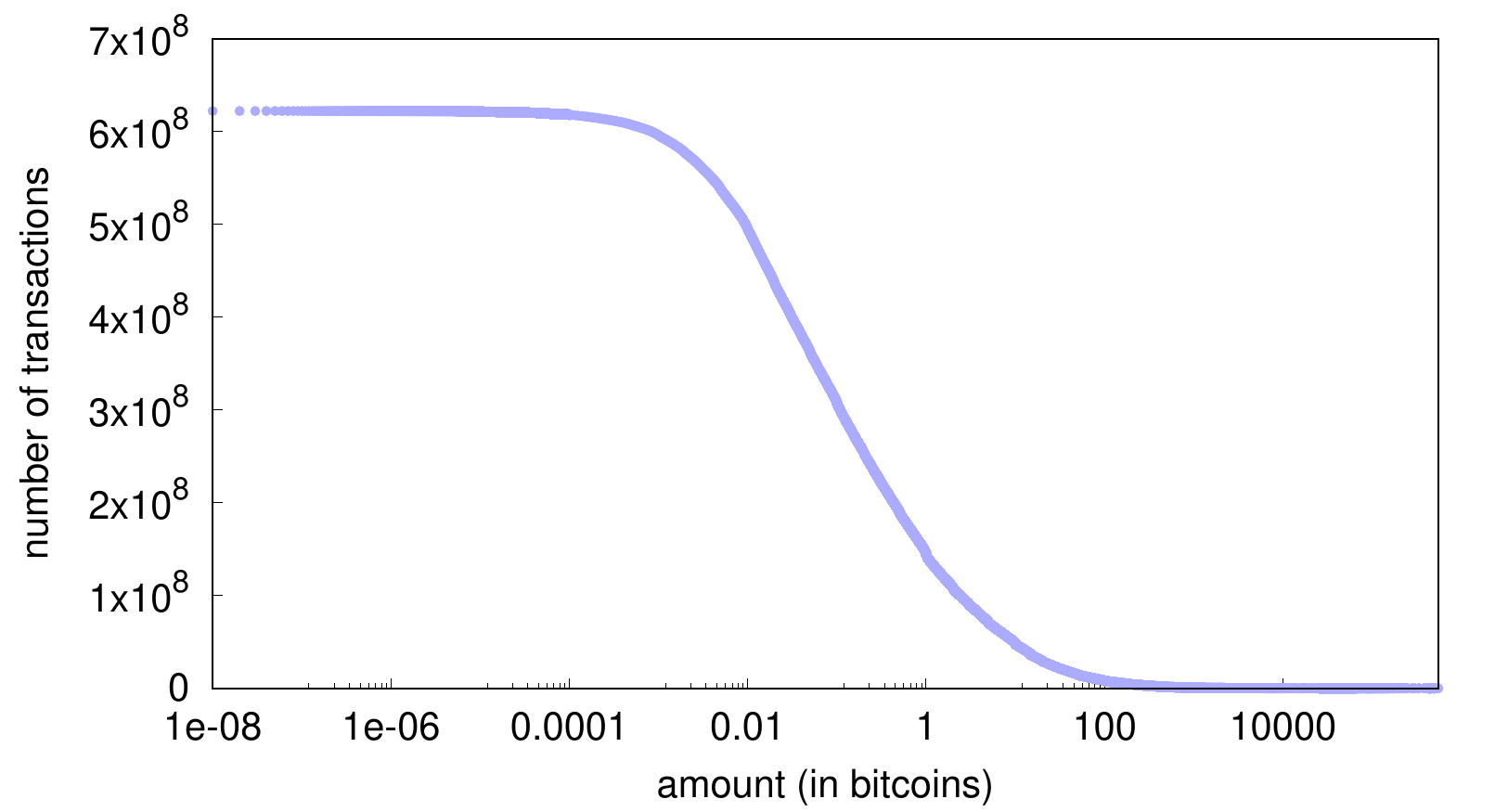}
\hfill
\includegraphics[width=.4\columnwidth]{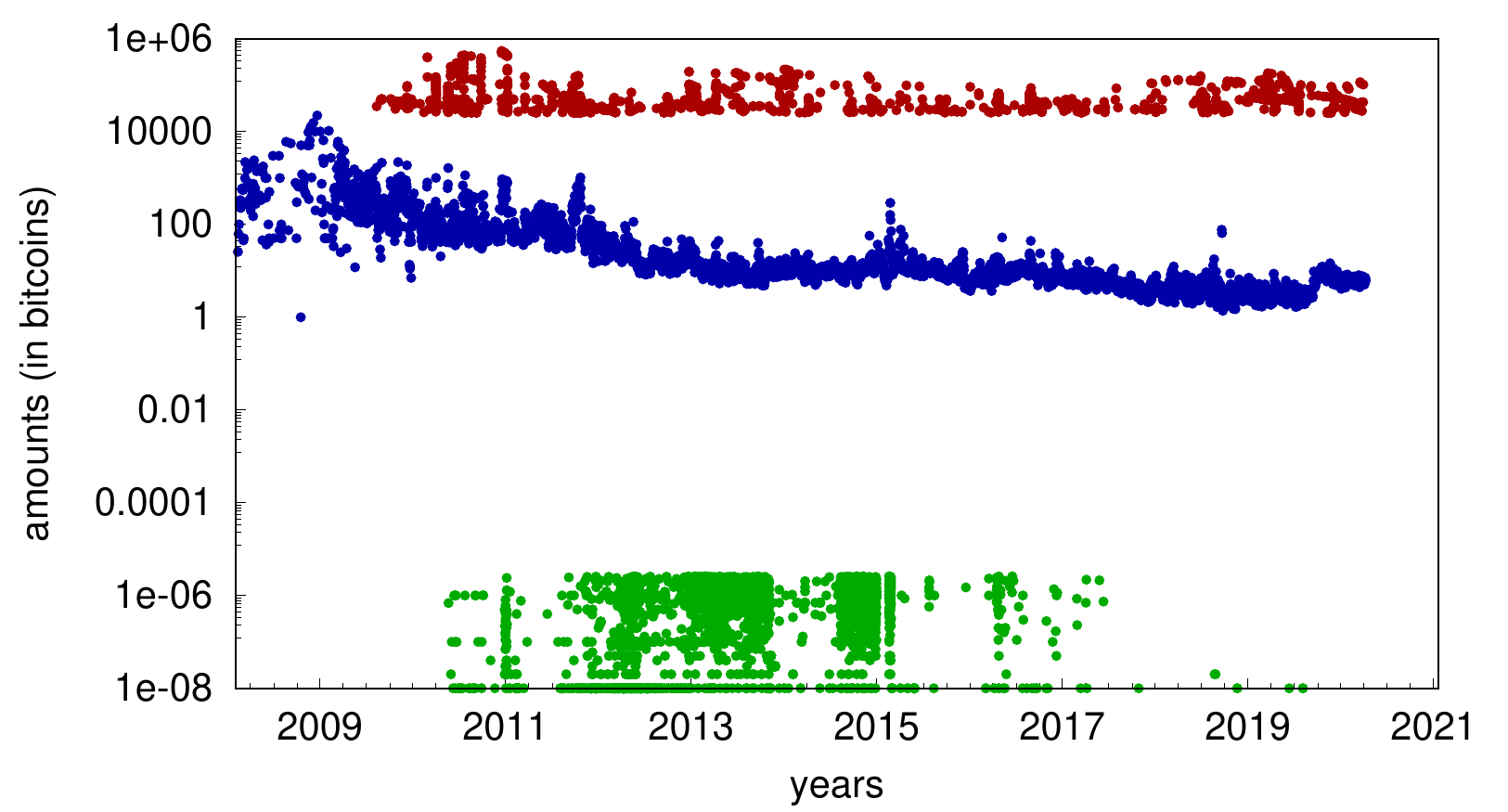}
\hfill~
\caption{
{\bf Left: inverse cumulative distribution of transaction amounts.} For each amount $x$ on the horizontal axis, we display the number of transactions with amount at least $x$ bitcoins (obtained as the sum of their output \field{value} fields), in log-lin scales.
{\bf Right: transaction amounts over time.} For each of the $10,000$ transactions with largest amounts (in red, top dots), and for each of the $10,000$ ones with smallest amounts (in green, bottom dots), we display a dot at its time occurrence (horizontal axis) and amount in bitcoins (vertical axis, in log scale). In between, we plot (in blue, middle dots) the daily average transaction amount.
}
\label{fig:trans_amount_t_and_amount_d}
\end{figure}

We also present in Figure~\ref{fig:trans_amount_t_and_amount_d}~(left) the transaction amount distribution, in bitcoins. These amounts span $11$ orders of magnitude, which is huge. Notice however that only $1$ transaction over $6$ has an amount of more than $1$ bitcoin. One may guess that large amounts are for old transactions, when bitcoin value was very low, and that more recent transactions have much lower amounts.  Figure~\ref{fig:trans_amount_t_and_amount_d}~(right) shows that this is not true: although the average amount significantly decreases, there are still many recent transactions with large amounts, and instead very small amounts tend to disappear. This certainly is a consequence of the generalization of platforms that group user-level transactions into large blockchain transactions for obfuscation and optimization reasons.

{\bf These plots provide good illustrations of what is readily feasible with the collected dataset.}
However, other key computations like counting the number of distinct addresses, or the number of occurrences of each address, are more subtle. They require a parsing of the whole data and storing the set of already seen addresses. Likewise, information on transaction inputs, like their associated addresses and satoshi amounts, is not readily available; finding them requires to store and query past transaction outputs. {\bf The next section is devoted to making such operations easy.}

\section{Indexing}
\label{sec:indexing}

Handling data at bitcoin blockchain scale requires appropriate indexing and pre-processing. {\bf Indexes from $0$ to $n-1$, where $n$ is the number of items of a given kind, are particularly appealing}: one may then store information related to item of index $i$ in an array of size $n$ and access it very efficiently (in both space and time).

For instance, with addresses indexed this way, it is easy to count their occurrences: from an array initially filled with $0$s, simply parse the data and increment the $i$-th cell each time address of index $i$ occurs. The time cost of this counting is dominated by the data parsing, since it has a small $O(1)$ cost per address occurrence. Its space cost is limited to an array of $n$ integers.

Another desirable feature is to have {\bf indexes consistent with occurrence order}: the first occurring item has index $0$, the next has index $1$, and so on.
With such indexings, counting address occurrences for the $n'$ first addresses, for $n'\le n$, requires an array of $n'$ integers only. Otherwise, an array of size $n$ is needed, like in the base case.
More generally, any prefix of the data has indexes from $0$ to $n'-1$, where $n' \le n$ is the number of items occurring in the prefix. This ensures that prefixes are themselves datasets encoded in the same way. We therefore say that {\bf such indexes are {\em prefix-consistent}}. There is a unique such indexing for a given dataset.

Here, items of interest are blocks, transactions, their inputs and outputs (TIOs), and addresses. As explained above, the \field{height} field of blocks already is its prefix-consistent index. Transactions have a \field{txid} field, that gives their native identifier (a 32~byte hash code). Similarly, addresses are alphanumeric character strings of variable length, representing public keys. The case of TIOs is more subtle: a TIO is uniquely identified by the \field{txid}\ of the transaction that created it (as an output), together with its rank in the output list of this transaction. Therefore, we consider the pair composed of this \field{txid} and this rank as the native identifier for this TIO. For instance, the first and second outputs of a transaction $X$ have native identifier $X,0$ and $X,1$, respectively. They may later appear as inputs of other transactions.

In order to prepare the collected data for more advanced analysis (Section~\ref{sec:application}), {\bf our goal here is to associate its prefix-consistent index to each occurrence of transaction, TIO, and address native identifier} in our dataset. Aiming at preserving its integrity, we however keep all the initial data and just add indexes within its \json\ objects, in dedicated fields. For instance, we add to each \json\ object having a \field{txid} field a new field, named \field{index} if this object is a transaction and named \field{txid\_index} if it is a transaction input, with value the prefix-consistent index of this transaction.

The {\bf classical approach} for prefix-consistent indexing is to store index tables in main memory, and then to perform indexing on-the-fly, while parsing the original data. However, this has prohibitive space and time costs.

Indeed, under reasonable assumptions, this approach requires to store at least the native identifiers in main memory. With the numbers of items above this leads to at least $60$~GB of memory needs.

In addition, there are more than $27.6$~billion item occurrences in total, and so we have to perform this number of searches in the indexes. The only data structure that fits the above space requirements are (sorted) arrays of native identifiers; but then searching has a prohibitive $\Theta(\log(n))$ time cost.
Hash tables make index operations very fast, but have space requirements significantly larger than necessary: in addition to native identifiers, they must store indexes, and some spare space. Moreover, these approaches make un-ordered accesses to huge memory spaces, which has a strong impact on speed in practice.

\begin{figure}[!h]
\centering
\includegraphics[width=.8\columnwidth]{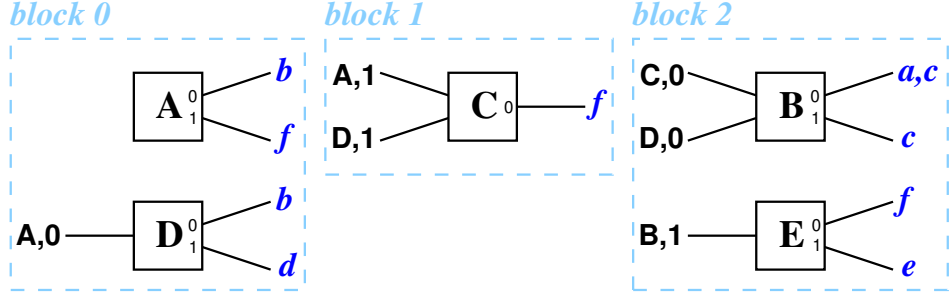}
\caption{{Example data for indexing.}
We consider three blocks, numbered 0, 1 and 2. Each contains transactions, named $A$, $B$, $C$, $D$, and $E$. Their input and output information are displayed as they are available in the dataset. We do not display amounts, timestamps, and other information that plays no role here.}
\label{fig:ex-index}
\end{figure}

Therefore, {\bf we propose an approach that avoids storing indexes in central memory and still parses the data a very limited number of times.} This approach relies on standard command-line tools, in particular \sort. It reads the raw data and adds index fields on-the-fly. To achieve this, we will build a file containing on its $i$-th line the pair \myline{i-1 index}, meaning that \myline{index} is the index of the $(i-1)$-th native identifier occurrence. Then, on-the-fly indexing is easy: it suffices to jointly parse the original data and this file. We detail the successive steps in the following, and illustrate intermediate results in the case of Figure~\ref{fig:ex-index}.

{\bf Our first step} consists in listing all occurrences of items under concern, together with their occurrence rank. In some cases, their index is readily available: when we parse the $i$-th transaction, we encounter its \field{txid} for the first time, and so its index is $i-1$; each TIO is created as a transaction output, therefore counting the number of transaction outputs encountered so far gives current TIO index. In these cases, we output the triplet \myline{occurrence-rank native-identifier index}. In other cases, when the index of current item is not directly available, we only output the pair \myline{occurrence-rank native-identifier}.

In the case of Figure~\ref{fig:ex-index}, this leads to the following:
\myline{0 A 0}
\myline{1 A,0 0}
\myline{2 b}
\myline{3 A,1 1}
\myline{4 f}
for transaction $A$,
\myline{5 D 1}
\myline{6 A}
\myline{7 A,0}
\myline{8 D,0 2}
\myline{9 b}
\myline{10 D,1 3}
\myline{11 d}
for transaction $D$,
\myline{12 C 2}
\myline{13 A}
\myline{14 A,1}
\myline{15 D}
\myline{16 D,1}
\myline{17 C,0 4}
\myline{18 f}
for transaction $C$,
\myline{19 B 3}
\myline{20 C}
\myline{21 C,0}
\myline{22 D}
\myline{23 D,0}
\myline{24 B,0 5}
\myline{25 a}
\myline{26 c}
\myline{27 B,1 6}
\myline{28 c}
for transaction $B$,
and
\myline{29 E 4}
\myline{30 B}
\myline{31 B,1}
\myline{32 E,0 7}
\myline{33 f}
\myline{34 E,1 8}
\myline{35 e}
for transaction $E$.

In the obtained list, the native identifier of each item appears exactly once in a triplet with their index, except the ones of addresses. {\bf Our second step} aims at building similar triplets for addresses. To do so, we parse the original data again and list all address occurrences together with their rank. We then sort this list with respect to addresses and we keep only the first tuple for each address. This gives the list of all distinct addresses, each with its first occurrence rank. We sort again with respect to this field, in order to obtain the list of all distinct addresses, ordered by rank. The index of any address is nothing but its rank in this list, and we finally output a triplet \myline{-1 address index} for each address.

In the case of Figure~\ref{fig:ex-index}, we first obtain
\myline{b 0}
\myline{f 1}
\myline{b 2}
\myline{d 3}
\myline{f 4}
\myline{a 5}
\myline{c 6}
\myline{c 7}
\myline{f 8} and
\myline{e 9}.
We then obtain
\myline{a 5}
\myline{b 0}
\myline{c 6}
\myline{d 3}
\myline{e 9} and
\myline{f 1},
then 
\myline{b 0}
\myline{f 1}
\myline{d 3}
\myline{a 5}
\myline{c 6} and
\myline{e 9}.
Finally we obtain
\myline{-1 b 0}
\myline{-1 f 1}
\myline{-1 d 2}
\myline{-1 a 3}
\myline{-1 c 4} and
\myline{-1 e 5}.

These triplets are in a format similar to the ones from our first step. In our {\bf third step} we put these two sets of tuples together, and sort them according to their second field, which is a native identifier, as well as their first field, numerically. This ensures that all tuples mentioning a given native identifier are grouped together. The first of them necessarily is a triplet, and its third field is the index of the native identifier under concern (second field).

In our example, we obtain
\myline{-1 a 3}
\myline{25 a}
\myline{0 A 0}
\myline{6 A}
\myline{13 A}
\myline{1 A,0 0}
\myline{7 A,0}
\myline{3 A,1 1}
\myline{14 A,1}
\myline{-1 b 0}
\myline{2 b}
\myline{9 b}
\myline{19 B 3}
\myline{30 B}
\myline{24 B,0 5}
\myline{27 B,1 6}
\myline{31 B,1}
\myline{-1 c 4}
\myline{26 c}
\myline{28 c}
\myline{12 C 2}
\myline{20 C}
\myline{17 C,0 4}
\myline{21 C,0}
\myline{-1 d 2}
\myline{11 d}
\myline{5 D 1}
\myline{15 D}
\myline{22 D}
\myline{8 D,0 2}
\myline{23 D,0}
\myline{10 D,1 3}
\myline{16 D,1}
\myline{-1 e 5}
\myline{35 e}
\myline{29 E 4}
\myline{32 E,0 7}
\myline{34 E,1 8}
\myline{-1 f 1}
\myline{4 f}
\myline{18 f}
\myline{33 f}.

This leads to the {\bf fourth and last step} of our indexing procedure. For each native identifier, we take its index on the first tuple mentioning it, and for each tuple mentioning it (except the ones starting with \myline{-1}), we output a pair \myline{i index}, where $i$ is the first field of the tuple. Such a pair says that the $i$-th item occurrence in the dataset corresponds to index \myline{index}. We therefore sort these pairs according to their first field, and obtain the wanted file.

In our example, this leads to
\myline{0 0}
\myline{1 0}
\myline{2 0}
\myline{3 1}
\myline{4 1}
\myline{5 1}
\myline{6 0}
\myline{7 0}
\myline{8 2}
\myline{9 0}
\myline{10 3}
\myline{11 2}
\myline{12 2}
\myline{13 0}
\myline{14 1}
\myline{15 1}
\myline{16 3}
\myline{17 4}
\myline{18 1}
\myline{19 3}
\myline{20 2}
\myline{21 4}
\myline{22 1}
\myline{23 2}
\myline{24 5}
\myline{25 3}
\myline{26 4}
\myline{27 6}
\myline{28 4}
\myline{29 4}
\myline{30 3}
\myline{31 6}
\myline{32 7}
\myline{33 1}
\myline{34 8}
\myline{35 5}.

As explained above, {\bf we finally perform the on-the-fly index addition} by parsing the original data and the obtained list of pairs jointly. In the whole procedure, no index is stored in central memory, and we never had to sort the original \json\ file. Instead, we sort the list of all item occurrences and the index by taking benefit from the highly optimized \sort\ tool, that uses external memory to handle huge tasks and only uses a user-specified amount of main memory. {\bf In our settings, the full indexing took approximately $98$~hours (close to $4$~days)}, which is very reasonable given the fact that the initialization of the bitcoin node took $6$~hours, and the \rpc\ \json\ data collection itself took $58$~hours, approximately.

Notice that using an advanced sorting tool as above is not mandatory. If only a routine sorting function is available, then one may proceed as follows. First divide the total number of item occurrences $M$ into parts of size $N$ such that sorting $N$ items fits in central memory. These $N$ occurrences involve $N' \le N$ distinct native identifiers. The procedure above is able to index them in the whole data. Running this $\frac{M}{N}$ makes the whole indexing, at the cost of repeated parsing of the original data. In this way, the whole procedure may be turned into a standalone program.

\section{Distillation}
\label{sec:distillation}

All operations above preserve the original data; they only add address and index fields in order to make it easier to use. However, in most practical situations, one needs specific parts of the data only. Then, systematically resorting to the global dataset is an overkill: the file is huge and it requires \json\ parsing which, in addition to a waste in computation time, requires a \json\ parsing library.

Instead, users need to easily and quickly run their computations, with controlled space needs, for instance using low-level languages like C. This requires a filtering and pre-processing of the dataset, called {\em distillation}. {\bf It extracts the needed information and puts it in a convenient format for easy, fast, and compact processing.} We illustrate such distillations in this section, as well as their practical uses.

First notice that, although relevant data depends on the targeted use, many need similar data. In the case of bitcoin, one is typically interested in the {\bf list of transactions with their timestamp, input addresses, and output addresses}. For this reason, we show how to distillate our dataset into a sequence of one line per transaction, each with the following space-separated fields:
\myline{block timestamp tx nb-in nb-out first-in ... last-in first-out ... last-out}. Here, \field{tx} stands for the index of the transaction under concern; \field{block} is the index of the block that contains this transaction, and \field{timestamp} is its timestamp; \field{nb-in} and \field{nb-out} give for the number of addresses involved in this transaction inputs and outputs, respectively; and the two sequences \field{first-in ... last-in} and \field{first-out ... last-out} give the indexes of these \field{nb-in} and \field{nb-out} addresses, respectively.

In the case of our guiding example (Figure~\ref{fig:ex-index}), the lines of distilled data are:
\myline{0 t0 0 0 2 0 1}
\myline{0 t0 1 1 2 0 0 2}
\myline{1 t1 2 2 1 1 2 1}
\myline{2 t2 3 2 2 0 1 3 4} and
\myline{2 t2 4 1 2 4 1 5}
where \field{t0}, \field{t1} and \field{t2} stand for the timestamps of blocks $0$, $1$, and $2$, respectively.

{\bf We perform this distillation as follows.} We parse the indexed dataset block by block, and store the current block index and timestamp. We parse transactions in current block in their order within the block, and consider current transaction index. This gives the three first fields of the output line for current transaction. We then parse its inputs and outputs and build the corresponding sets of addresses. Output addresses are directly available within the current transaction \json\ object. Instead, input addresses are not readily available. But transaction inputs are always outputs of {\em former} transactions. We therefore store the addresses of each encountered transaction output, in an array indexed by TIO indexes. When a TIO is encountered as another transaction input, we query this array and obtain the corresponding addresses, which gives us all needed information.

 Notice that transaction outputs are used only once as other transaction inputs. Therefore, keeping the addresses of already encountered transaction inputs is unnecessary. We provide an implementation of this procedure with this space optimization (we drop transaction output addresses once used as input) \cite{URL}.

Our implementation of this procedure performed the whole distillation in $25$~hours, leading to a $12$~GB file containing the key information listed above (the mere parsing of the \json\ file already takes $15$~hours). During this process, like in most distillation tasks, {\bf the indexes produced in previous section play a crucial role}. They make it possible to handle data in a very fast and compact way, with arrays queried by item indexes.

{\bf The distilled dataset is easy to parse}, even in a low-level language like C or with shell scripts. Indexes make it easy, fast and compact to perform many operations. For instance, we obtain that $3.5$~\% of all addresses are used only once, $87.6$~\% only twice, and almost $1.2$~\% are used as input and output of a same transaction (often more than once). This reflects two well-known but rarely quantified bitcoin facts: many users collect transaction change on an address they already use; and many users avoid re-using addresses, for privacy concerns.

\begin{figure}[!h]
~\hfill
\includegraphics[width=.4\columnwidth]{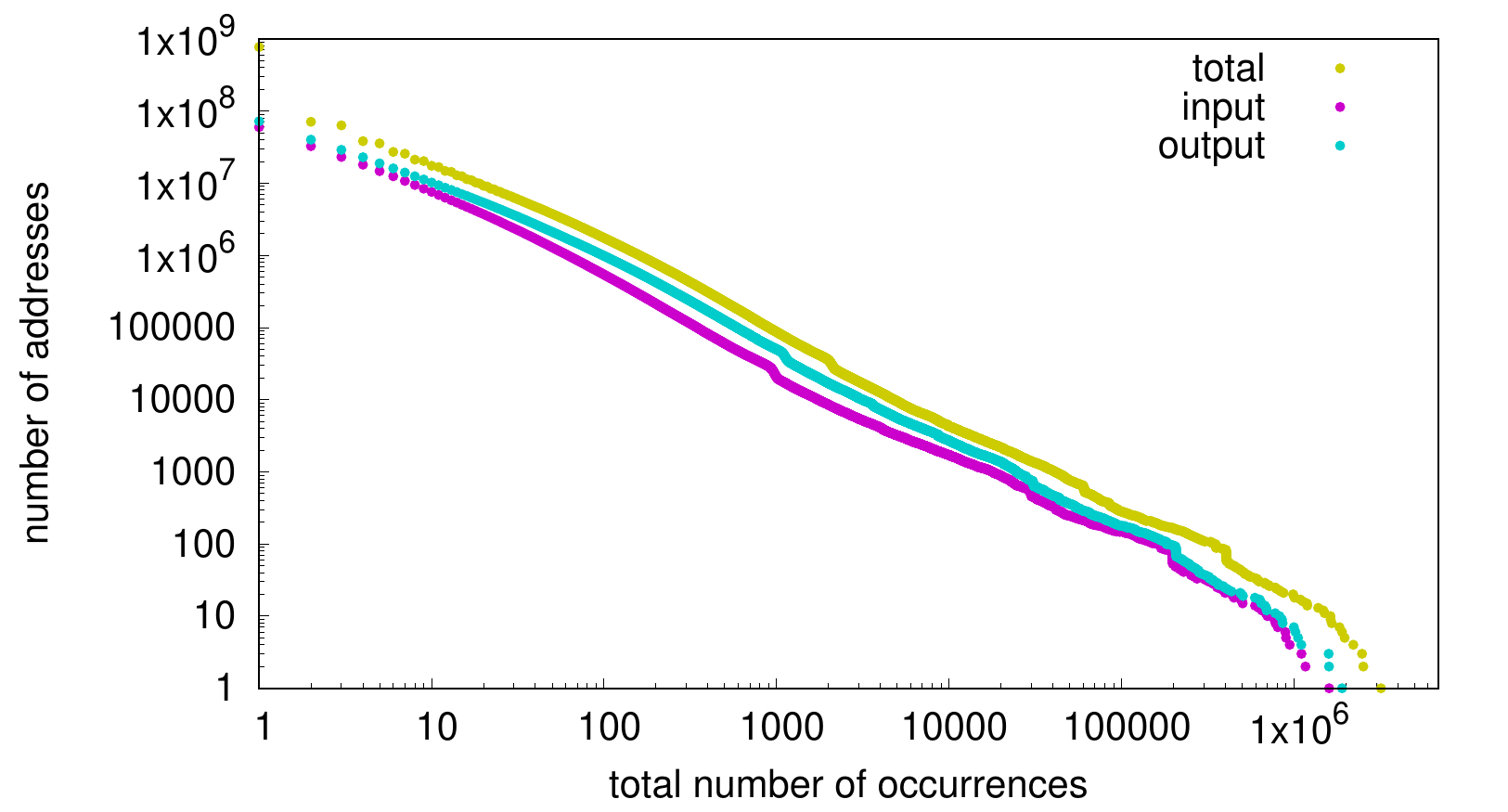}%
\hfill
\includegraphics[width=.4\columnwidth]{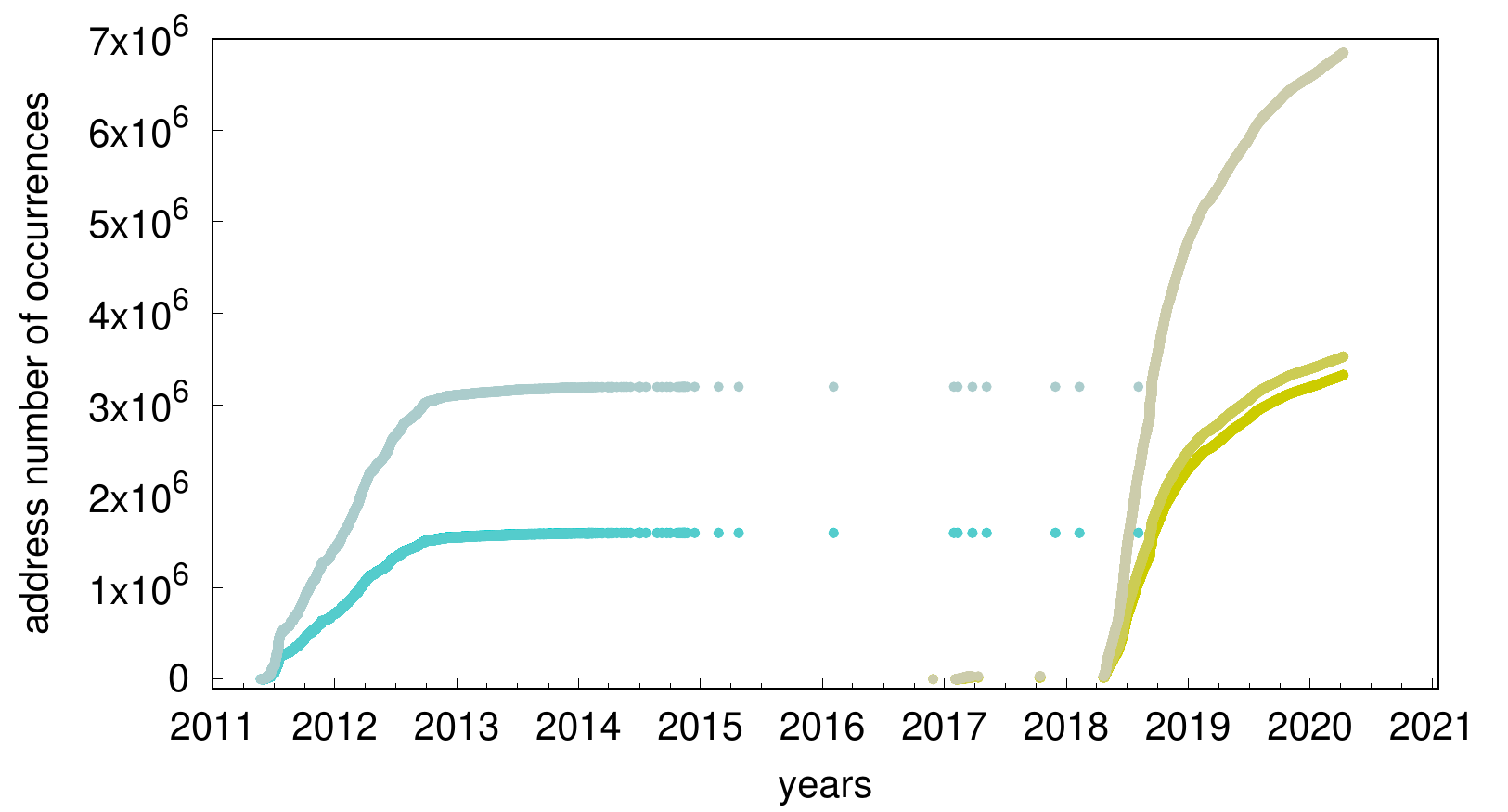}
\hfill~
\caption{
{\bf Left: address reuse}, displayed as the inverse cumulative distribution of the number of occurrences of each address in the dataset, in total, as an input address, or as an output address.
{\bf Right: occurrences of the two most used addresses over time.} For the most used address in the dataset (rightmost plots, green colors) and the second most used one (leftmost plots, blue colors), we display its number of occurrences as input, output, and its total number of occurrences since the beginning of the blockchain.
}
\label{fig:addr_reuse_d_and_io_addresses_d}
\end{figure}

We also obtain that the most frequent address in transaction inputs and outputs appear there $3,324,680$ and $3,525,298$ times, respectively.
More generally, we display in Figure~\ref{fig:addr_reuse_d_and_io_addresses_d}~(left) the distribution of the total number of occurrences of each address, as well as their numbers of occurrences as transaction inputs and outputs. All these distributions are very similar, and span more than $6$ orders of magnitude, showing a huge heterogeneity between addresses.

We display in Figure~\ref{fig:addr_reuse_d_and_io_addresses_d}~(right) the timeline of transactions involving the two most frequent addresses in our dataset. The first one appears as soon as 2011, and it is intensively used for one year. Its uses then severely slows down, and becomes sporadic after 2015, but, surprisingly enough, it is still used by the end of 2018. Its number of occurrences as input and output are so close to each other that we cannot distinguish them in the plot. The behaviour of the other address is quite different. It appears much later, just before 2017, but it is used only a few times until 2018. Then, its use is intensive for a few months. It slows down in 2019, and suddenly stops being used in mid-2020.

Many other statistics are easy to obtain from various distilled versions of our dataset.

\begin{figure}[!h]
~\hfill
\includegraphics[width=.4\columnwidth]{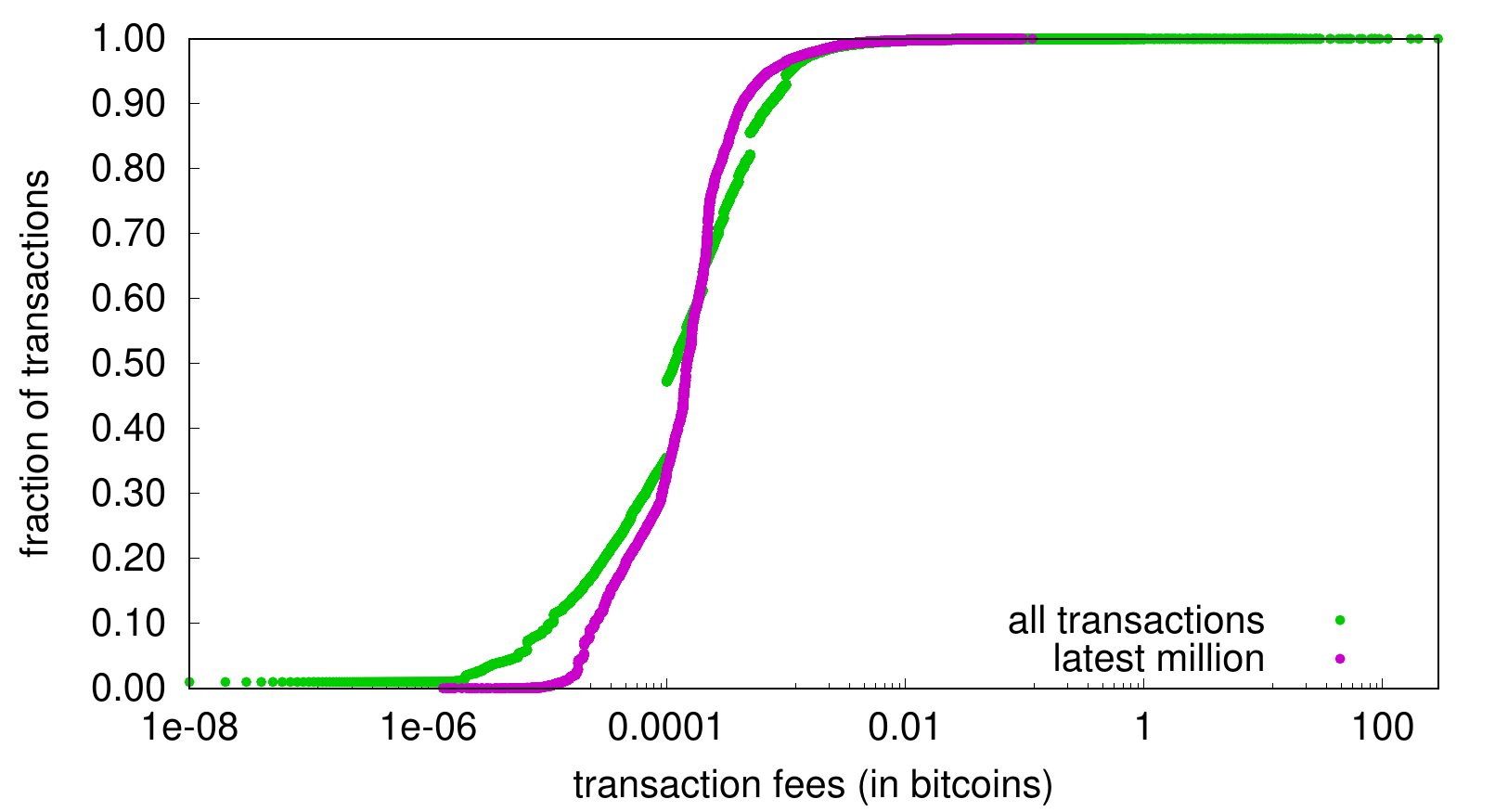}%
\hfill
\includegraphics[width=.4\columnwidth]{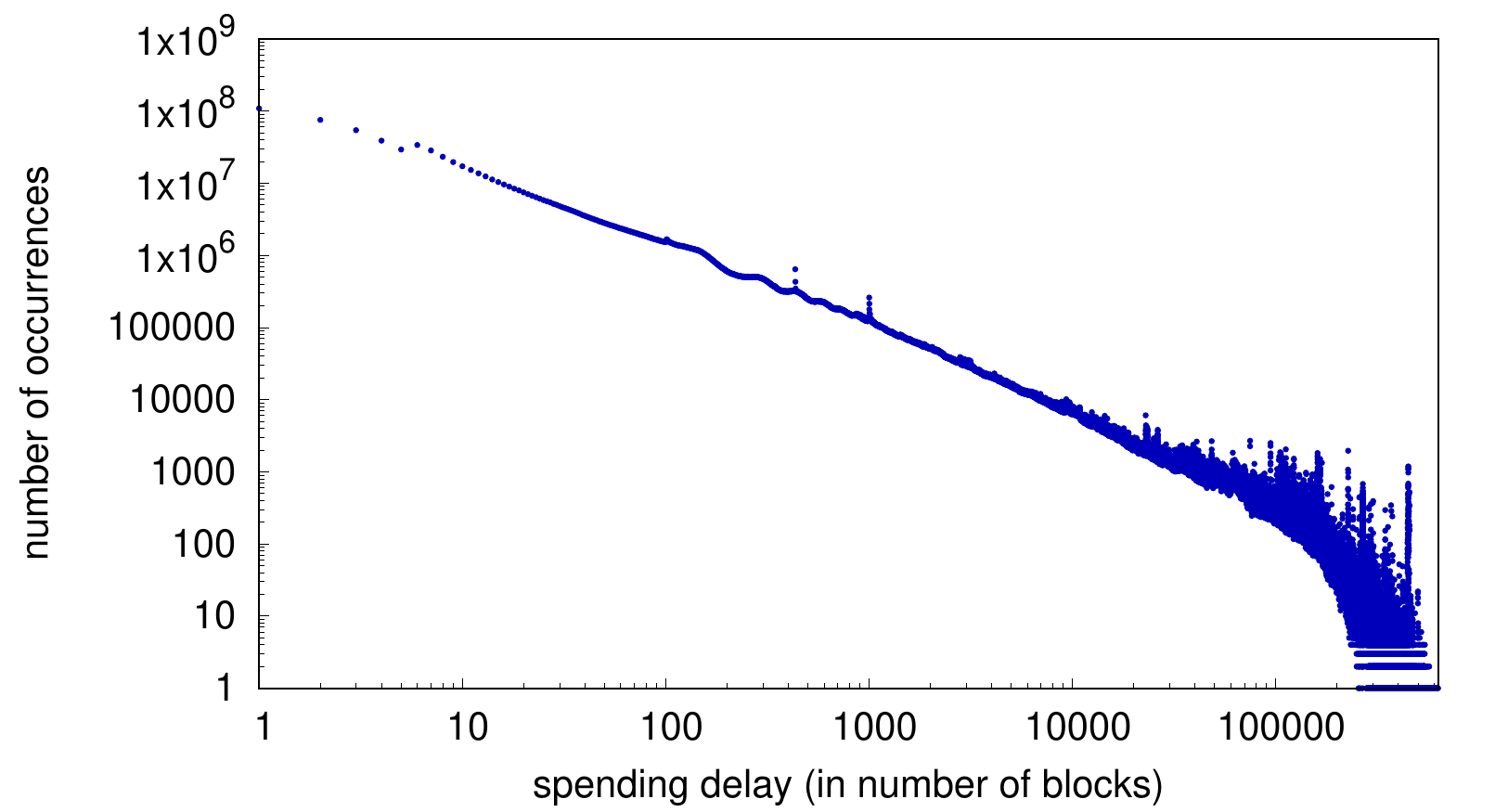}
\hfill~
\caption{
{\bf Left: transaction fees}, {\em i.e.} difference between transaction output and input amounts (in bitcoins) for the whole dataset and for the latest million transactions. We display the cumulative distributions: for each amount $x$ on the horizontal axis, we display a point indicating the fraction of transactions with fees (vertical axis) lower than this value.
{\bf Right: spending delay} distribution (log-log scales). The delay is the number of blocks issued between the time at which bitcoins are received as a transaction output, and the time at which they are spend as a transaction input.
}
\label{fig:fees_and_delays}
\end{figure}

For instance, replacing addresses by amounts, gives transaction fees (the difference between input and output amounts), displayed in Figure~\ref{fig:fees_and_delays} (left). Since the beginning of the blockchain, fees range from $0$ to more than $100$ bitcoins, but they are nowadays much more uniform: they range from $10^{-6}$ to $0.02$ bitcoins for the last million of transactions, with the vast majority very close to $0.0002$ bitcoins.

Replacing addresses by TIOs gives information on bitcoin flows, like the delay between the receiving and spending of bitcoins, displayed in Figure~\ref{fig:fees_and_delays} (right). Many bitcoins are spent at a fast pace (just a few blocks), and the delay distribution is very close to a power-law for 4 decades. However, it has a cut-off due to the limited number of blocks currently in the blockchain. Also, specific spending delays are over-represented, like the ones around $1000$ blocks.

{\bf Such statistics are easy to obtain with our dataset and methodology}. All computations above take a few dozens of minutes from the distilled dataset and need a few GB of central memory only. We present a more advanced application in next section.

\section{Application: address clustering}
\label{sec:application}

Transactions discussed above link input addresses to output addresses, but it often makes more sense to study bitcoin flows between individuals, companies, institutions, and other social entities. To do so, one has to {\bf build clusters of addresses} that arguably belong to the same entity, and then observe transactions between these address clusters.

Many heuristics exist to build such clusters \cite{DBLP:conf/complexnetworks/RemyRM17}. The most classical one is the {\em input-based heuristic}; it assumes that all input addresses of a transaction belong to the same cluster. Indeed, this transaction accesses the bitcoins associated to these addresses, and so it is reasonable to assume that a same entity owns them. We focus on this heuristic here for illustration purpose, but all the following applies the other heuristics.

Address clustering heuristics are mitigated by the use of various obfuscation and optimization methods \cite{10.1145/3205230.3205234}. Several of them, like for instance mixing, induce transactions with many and unrelated input addresses. Such transactions not only mistaken the input-based heuristic by erroneously indicating that their input addresses are related; they merge the clusters legitimately containing these addresses and so may have a dramatic impact on obtained clusters.

{\bf In order to illustrate the use of our dataset and the power brought by its pre-processing, we show here how to explore the relevance of this heuristic, and the impact of such transactions on its results.}

To do so, we consider two key metrics: the number of obtained clusters and the size of the largest one. We compute these metrics when one considers only transactions with at most \Kin\ inputs and \Kout\  outputs. Observing how they vary with \Kin\ and \Kout\ shows the impact of suspicious transactions (the ones with large numbers of inputs and/or outputs) on obtained clusters. Indeed, unwanted transactions merge clusters of addresses, and so they increase their size and decrease their number.

Cluster computations usually rely on a graph in which nodes are addresses and two addresses are linked if they appear as input of a same transaction. Then, the wanted clusters are the connected components of this graph. However, not all links are needed: it is sufficient to ensure that all input addresses of each transaction are reachable from each other in the graph. For instance, linking the first input address of a transaction to all its other input addresses is sufficient; linking the first to the second, the second to the third, and so on, also works. Choosing between all appropriate solutions may have a marginal impact on the speed of cluster computations, but in all cases $\kin - 1$ links are sufficient, for any transaction with \kin\ input addresses.

Building this graph may be done by parsing the distilled data and writing as output the wanted links between address identifiers, $\kin-1$ per transaction with $\kin$ outputs. Then, these links are sorted to remove duplicates, and the graph is loaded into central memory to compute its connected component, in time and space linear with its number of links.

However, {\bf a union-find approach} \cite{cormen} is faster, more compact, and simpler. It uses the same list of links, but it does not need to sort it, and it only stores in central memory an array of all input address identifiers. It then needs a constant amortized time per link in the list to {\em find} the components of its nodes and perform their {\em union} if needed. The overall time cost therefore is linear in the number of links in the list, which itself is linear in the total number of transaction inputs. Our address indexes are numbers from $0$ to $n-1$ where $n$ is the total number of addresses. Therefore, this method only needs to store $n$ integers (of value at most $n$) in central memory.

Notice however that we investigate here the impact on obtained clusters of the considered maximal numbers \Kin\ and \Kout\ of input and output addresses. We therefore need to compute clusters for various values of these parameters. In order to reduce the redundancy of these computations, we perform them for all values of \Kin\ in a row: we sort the link list according to the number of input addresses of the corresponding transaction, and output obtained information when the number of input addresses grows. We repeat this for all meaningful value \Kout\ of the maximal number of output addresses.

{\bf In summary, we proceed as follows}: we parse the distilled data and consider each transaction; for its \kin\ inputs, we list the $\kin - 1$ links between the index of the first input address and all the others,
together with \kin\ and its number of outputs \kout; we sort this list according to the \kin\ field in order to have all links in increasing transaction input number; then, for any given maximal number of output addresses \Kout\, we perform union-find based on this sorted list by skipping lines with $\kout>\Kout$ and, whenever the \kin\ field grows, we output \kin, the current number of clusters, and the current maximal cluster size. This gives the wanted metrics as a function of the maximal number of input addresses \Kin, for the considered maximal number of output addresses \Kout.
Our implementation \cite{URL} obtains the results in a few dozens of minutes only.


\begin{figure}[!h]
~\hfill
\includegraphics[width=.4\columnwidth]{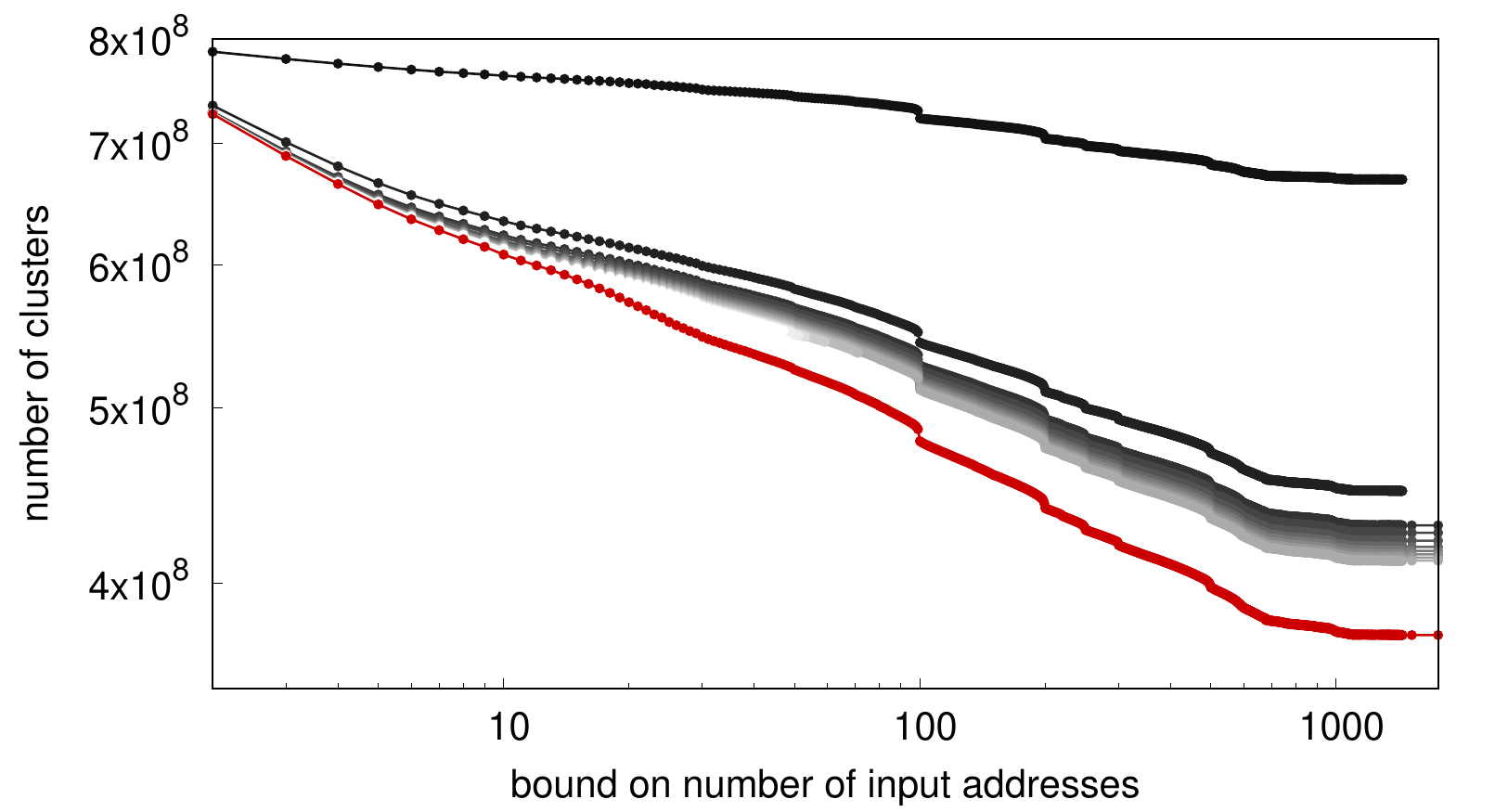}%
\hfill
\includegraphics[width=.4\columnwidth]{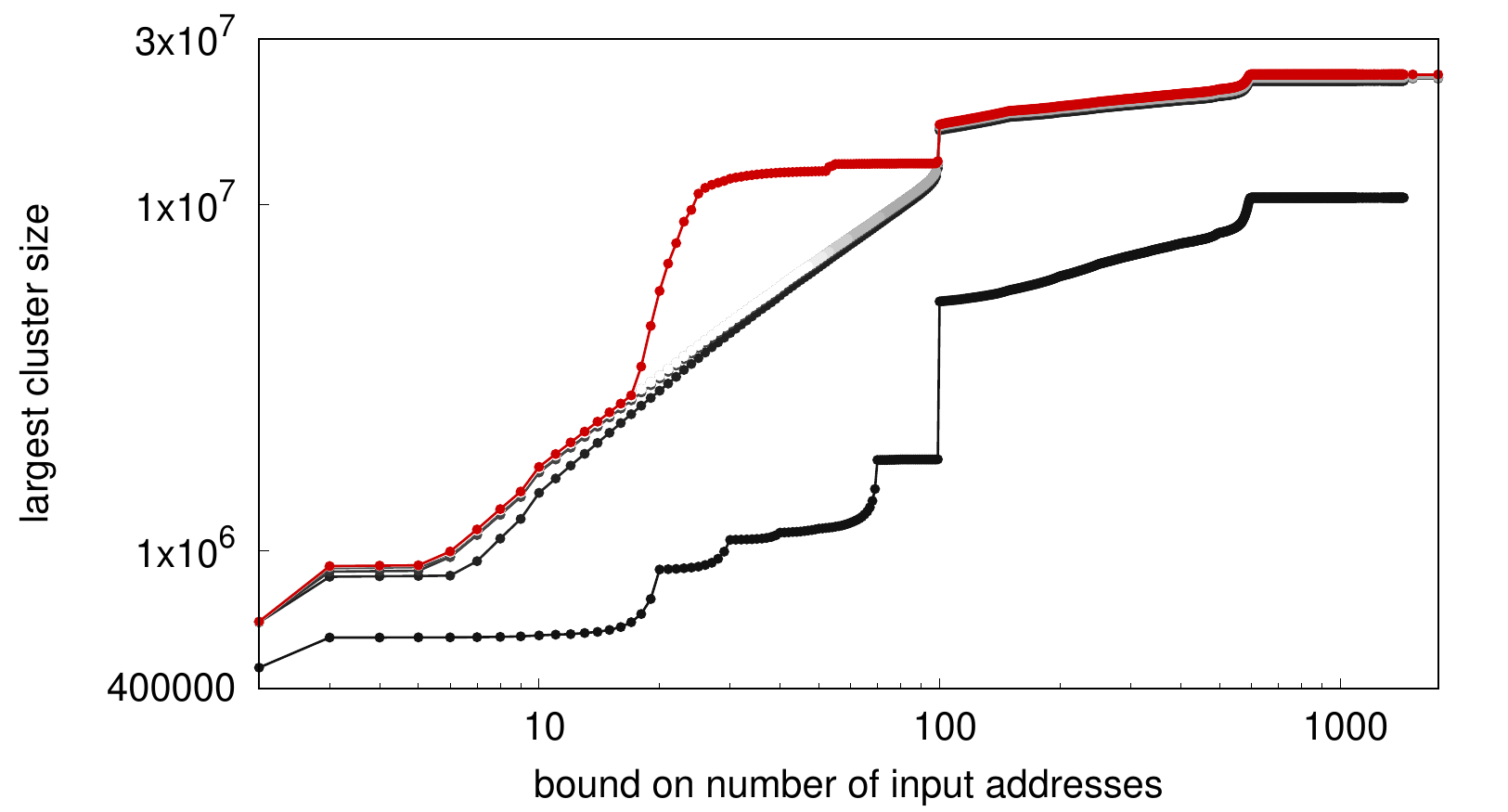}
\hfill~
\caption{
{\bf Cluster metrics:} left: number of clusters; right: maximal cluster size. Each metric is displayed in log scale on the vertical axis, as a function of the maximal number \Kin\ of input addresses of considered transactions, on the horizontal axis (log scale). We display the plots obtained for each value of \Kout\ from $1$ to $15$, as well as the one obtained with no limitation on the number of outputs.
The number of clusters (left) decreases with the number of considered transactions, therefore the bottommost plot (in red) is the one for any output number. Conversely, the maximal cluster size (right) increases with the number of considered transactions, therefore the topmost plot (in red) is the one for any output number.
}
\label{fig:cluster_metrics}
\end{figure}

Figure~\ref{fig:cluster_metrics} displays the obtained plots. As expected, they show that using transactions with larger numbers of inputs and/or outputs has a strong impact on detected clusters (both their size and number). There is a clear difference between the partitions into clusters obtained when the maximal number of inputs \Kin\ if below $10$ or $20$, and the ones obtained for larger values of \Kin. Likewise, although we display plots for values of \Kout\ up to $15$, they are still very different from the ones obtained with no limit on the number of outputs.

Deepening these observations is out of the scope of this paper, but we can already conclude that transactions with more than a dozen inputs or outputs have a strong impact on the input-based heuristic; they certainly play a specific role and should be handled with care. To go further, one may observe cluster size distributions in the same way, investigate other heuristics, and refine them. For instance one may consider weighted graphs between addresses, where the weight may represent the number of times addresses co-occur as input addresses, or other measures of heuristics reliability. This would not significantly change the procedure above.

Up to our knowledge, this is the first time that such plots are observed. {\bf Certainly, the lack of available and indexed data was the limiting factor for such investigations.} As illustrated above, our dataset makes them much easier and faster to obtain than before.

\section{Related work}
\label{sec:related}

Bitcoin is at the core of an intense research activity.
A comprehensive review being out of reach, we focus here on data collection and management, with an empĥasis on a few key contributions.

{\bf Many companies store their own copy of the bitcoin blockchain and provide partial access to it} through dedicated web forms, APIs and/or CSV files. Their data is stored in local databases, and they provide additional information like charts, top users, address clusters, mixer lists, etc. Queries based on \field{txid}s and addresses are in general possible, as well as browsing between transactions.

For instance, some focus on user-friendly browsing \cite{BlockStreamURL,blockchainCompanyURL}, with {\em BlockStream} providing open-source tools. The {\em Sphere 10 Software} company proposes a SQL-based approach \cite{SphereURL}. {\em WalletExplorer} and {\em Chainalysis} emphasize identification of services, mostly based on manual investigation \cite{WalletExplorerURL,ChainalysisURL}. {\em Blockchair} handles a wide variety of crypto-currencies \cite{BlockchairURL}, {\em OXT} has a particularly ideological policy \cite{OXTURL}, while {\em Kaiko} targets business applications \cite{KaikoURL}.

Such data providers have two important strengths: a user-friendly access to bitcoin data, and additional information to help analysis.
However, they provide very partial views: they focus on some aspects of the data, and large-scale computations are not efficient through such interfaces. Therefore, {\bf researchers often conduct their own data collection, and provide the obtained datasets.}

The most advanced such dataset \cite{BitcoinTransactionNetworkURL} probably is the one provided by Kondor et al \cite{Kondor2013,Kondor2014}. The authors patch the bitcoin software in order to record transaction information. Then, they compute identifiers similar to ours, and perform address clustering. They provide data collection and analysis tool source code, as well as the data up to February 2020.

{\bf Other authors design and provide data collection and management frameworks for bitcoin blockchain.}

An important contribution is the {\em Abe} free blockchain browser \cite{AbeURL}, that decodes binary blockchain files, stores data into a database, and provides query services. Others use Neo4j graph databases, like \cite{Neo4jBitcoinBin} that collects data by parsing the binary files, or \cite{Neo4jBitcoinRPC} that uses parallel \rpc\ calls, like us.
Likewise, \cite{general2017} proposes an open source Scala library using a local database \cite{general2017} for efficient high-level analytics. Some frameworks target specific challenges, like graph analytics in \cite{Towards2018,BitcoinTowardsURL}, that use a cluster for data collection (custom C++ bitcoin data decoding) and analysis (Neo4j database). More recently, \cite{Scalable2021} target user and transaction profiling, using a Spark-based tool.

These frameworks based on general-purpose databases are outperformed by dedicated, highly optimized in-memory approaches. In particular, {\em BlockSci} \cite{BlockSci} provides a very efficient infrastructure to handle bitcoin (and other) blockchain data. It stores them in central memory and collects it using binary file decoding and \rpc. It is used for instance in \cite{sharma2020bitcoins}.
{\em DataChain} \cite{DataChain2019} is another example; it provides a lightweight, flexible and interoperable framework for high-level queries to blockchains.

Most of these works provide source code of their collection and analysis tools, and/or the datasets they obtain. It must be clear however that, because of the continuous evolution of bitcoin technologies, maintaining up-to-date tools and datasets is challenging; {\bf many of them are outdated}. This is particularly true for methods that decode the bitcoin binary files.

In addition, {\bf these tools have non-trivial requirements}: some need to compile complex tools, many rely on local database systems that require terabytes of disk space, others have huge central memory needs, etc.

Last but not least, to the best of our knowledge, {\bf no available dataset provides the full information available in bitcoin blockchain}: they focus on parts of the data and only decode these parts; and/or they assume that some information is irrelevant and discard it. Even the Kondor et al dataset \cite{BitcoinTransactionNetworkURL}, which seems to be the most complete one, does not contain all blockchain data. In addition, such datasets often result from quite complex and poorly documented pre-processing steps, like data cleaning or clustering, and only the results of these pre-processings are provided. This may help for some dataset uses, but this may be a limitation for others; in all cases, this raises reproducibility and interpretation concerns.



\section{Conclusion and perspectives}
\label{sec:discussion}

{\bf Our work provides a comprehensive view of bitcoin blockchain by capturing absolutely all data it contains.}
In contrast with previous works, it relies only on the simplest and most standard tools. 
 It only requires a few hundred GB of disk space, including indexing. It is then easy to obtain very compact extractions of interest, and the indexes make them very convenient for a wide variety of studies. We give a thorough description of our procedure, with fully detailed application cases and documented code \cite{URL}.


Our collection procedure may be improved in several ways. The following are particularly appealing: a dataset updating function, ideally in real-time; a parallel \rpc\ procedure to reduce data collection time; querying pre-existing bitcoin nodes to avoid setting up our own node; as well as more advanced decoding of addresses and further analysis of ill-formed/non-standard transactions.

The {\bf data pre-processing may also be improved}. One may split the dataset into several (compressed) files, and store the position of blocks or transactions (or other entities of interest) in these files. Browsing the dataset in non-sequential ways would then be easy, while keeping the representation compact. Further improvements may use database structures like B-trees, or even database libraries. This would bring our contribution closer to database-oriented solutions, though, with their advantages and drawbacks.

\medskip

\noindent
{\footnotesize\em
{\bf Acknowledgements.} This work is funded in part by the ANR (French National Agency of Research) under the FiT LabCom grant.
}
\bibliographystyle{plain}
\bibliography{references}

\begin{thebibliography}{10}

\bibitem{AbeURL}
{\em Abe} free blockchain browser.
\newblock \url{https://github.com/bitcoin-abe/bitcoin-abe}.

\bibitem{blockchainCompanyURL}
{\em Blockchain company}.
\newblock \url{https://www.blockchain.com/explorer}.

\bibitem{BlockchairURL}
{\em Blockchair}.
\newblock \url{https://blockchair.com}.

\bibitem{BlockStreamURL}
{\em BlockStream}.
\newblock \url{https://blockstream.info/}.

\bibitem{ChainalysisURL}
{\em Chainalysis}.
\newblock \url{https://www.chainalysis.com/}.

\bibitem{Neo4jBitcoinBin}
How to import the bitcoin blockchain into {Neo4j}.
\newblock \url{https://neo4j.com/blog/import-bitcoin-blockchain-neo4j}.

\bibitem{Neo4jBitcoinRPC}
How to load bitcoin into neo4j in one day.
\newblock \url{https://medium.com/tokenanalyst}.

\bibitem{KaikoURL}
{\em Kaiko}.
\newblock \url{https://www.kaiko.com}.

\bibitem{OXTURL}
{\em OXT (Other/Open eXploration Tool)}.
\newblock \url{https://oxt.me}.

\bibitem{SphereURL}
{\em Sphere 10 Software}.
\newblock \url{http://blockchainsql.io}.

\bibitem{URL}
Supplementary material.
\newblock \url{http://bitcoin.complexnetworks.fr}.

\bibitem{WalletExplorerURL}
{\em WalletExplorer}.
\newblock \url{https://www.walletexplorer.com}.

\bibitem{general2017}
Massimo Bartoletti, Stefano Lande, Livio Pompianu, and Andrea Bracciali.
\newblock A general framework for blockchain analytics.
\newblock In {\em First Workshop on Scalable and Resilient Infrastructures for
  Distributed Ledgers}, 2017.

\bibitem{DBLP:conf/complexnetworks/RemyRM17}
R{\'{e}}my Cazabet, Rym Baccour, and Matthieu Latapy.
\newblock Tracking bitcoin users activity using community detection on a
  network of weak signals.
\newblock In {\em 6th Conference on Complex Networks and Their Applications},
  2017.

\bibitem{getblock-rpc}
Bitcoin Community.
\newblock bitcoin-core getblock rpc.
\newblock \url{https://bitcoincore.org/en/doc/0.21.0/rpc/blockchain/getblock}.

\bibitem{bitcoin-core}
Bitcoin Community.
\newblock bitcoin-core-0.21.0.
\newblock 2020.
\newblock \url{https://bitcoincore.org/bin/bitcoin-core-0.21.0}.

\bibitem{cormen}
Thomas~H. Cormen, Charles~E. Leiserson, Ronald~L. Rivest, and Clifford Stein.
\newblock {\em Introduction to Algorithms}.
\newblock The MIT Press.

\bibitem{10.1145/3205230.3205234}
Younggee Hong, Hyunsoo Kwon, Jihwan Lee, and Junbeom Hur.
\newblock A practical de-mixing algorithm for bitcoin mixing services.
\newblock In {\em 2nd ACM Workshop on Blockchains, Cryptocurrencies, and
  Contracts (BCC)}, 2018.

\bibitem{BlockSci}
Harry Kalodner, Malte M{\"o}ser, Kevin Lee, Steven Goldfeder, Martin Plattner,
  Alishah Chator, and Arvind Narayanan.
\newblock Blocksci: Design and applications of a blockchain analysis platform.
\newblock In {\em 29th {USENIX} Security Symposium}, 2020.

\bibitem{Kondor2014}
D{\'{a}}niel Kondor, Istv{\'{a}}n Csabai, J{\'{a}}nos Szüle, M{\'{a}}rton
  P{\'{o}}sfai, and G{\'{a}}bor Vattay.
\newblock Inferring the interplay between network structure and market effects
  in bitcoin.
\newblock {\em New Journal of Physics}, 2014.

\bibitem{Kondor2013}
D\'aniel Kondor, M\'arton P\'osfai, Istv\'an Csabai, and G\'abor Vattay.
\newblock Do the rich get richer? an empirical analysis of the bitcoin
  transaction network.
\newblock {\em PLoS ONE 9(2)}, 2014.

\bibitem{BitcoinTransactionNetworkURL}
D\'aniel Kondor, M\'arton P\'osfai, Istv\'an Csabai, and G\'abor Vattay.
\newblock {\em Bitcoin Transaction Network}, dryad, dataset.
\newblock \url{https://doi.org/10.5061/dryad.qz612jmcf}, 2021.

\bibitem{BitcoinTowardsURL}
Dan McGinn, Douglas McIlwraith, and Yike Guo.
\newblock Data from: Towards open data blockchain analytics: a bitcoin
  perspective, dryad, dataset.
\newblock \url{https://doi.org/10.5061/dryad.h9r0p65}, 2018.

\bibitem{Towards2018}
Dan McGinn, Douglas McIlwraith, and Yike Guo.
\newblock Towards open data blockchain analytics: a bitcoin perspective.
\newblock {\em Royal Society Open Science}, 2018.

\bibitem{whitepaper}
Satoshi Nakamoto.
\newblock Bitcoin: A peer-to-peer electronic cash system.
\newblock 2008.
\newblock \url{https://bitcoin.org/bitcoin.pdf}.

\bibitem{Scalable2021}
Raj~Sanjay Shah, Ashutosh Bhatia, Atith Gandhi, and Shray Mathur.
\newblock Bitcoin data analytics: Scalable techniques for transaction
  clustering and embedding generation.
\newblock In {\em 13th IEEE {COMSNETS}}, 2021.

\bibitem{sharma2020bitcoins}
Aman Sharma and Ashutosh Bhatia.
\newblock Bitcoin's blockchain data analytics: A graph theoretic perspective,
  2020.

\bibitem{DataChain2019}
Demetris Trihinas.
\newblock Datachain: A query framework for blockchains.
\newblock In {\em 11th ACM International Conference on Management of Digital
  EcoSystems}, 2019.

\end{thebibliography}
\end{document}